\documentclass{aa}
\usepackage{natbib}
\bibpunct{(}{)}{;}{a}{}{,} % to follow the A&A style

\usepackage{xcolor}
\usepackage{txfonts}
\usepackage{graphicx}
\usepackage{epsfig}
\usepackage{natbib}

\usepackage{url}
\usepackage{twoopt}
\usepackage{dcolumn}% Align table columns on decimal point
\usepackage[colorlinks=true]{hyperref}
\hypersetup{citecolor=blue}
\begin{document} 

\title{Asymmetrical shape of  the Near-Infrared K 4p-5s
        doublet perturbed by He and H$_2$}

\subtitle{}

\author{N. F. Allard         \inst{1,2}
   \and J. F. Kielkopf        \inst{3}
}

\institute{LIRA, Observatoire de Paris, Universit\'e PSL, Sorbonne
     Universit\'e, Sorbonne Paris Cit\'e, CNRS,
     61, Avenue de l’Observatoire, F-75014 Paris, France\\
              \email{nicole.allard@obspm.fr}
         \and
         Institut d'Astrophysique de Paris,  UMR7095, CNRS, 
         Universit\'e Paris VI, 98bis Boulevard Arago, F-75014 PARIS, France \\
          \and
          Department of Physics and Astronomy, 
          University of Louisville, Louisville, Kentucky 40292 USA \\  
}

\date{Received 3 June/ Accepted 25 July 2025}

\abstract{
 The goal of this  paper is to provide an exhaustive  study of the K $4p$--$5s$
 doublet at
1.243/1.252~$\mu$m perturbed by He and  H$_2$ in physical conditions of brown dwarfs.  This work is applicable to analysis of the
ESO SupJup Survey's J-band
spectra such as for Luhman~16.
Line profiles of this doublet  computed within a unified line-shape
  theory exhibit a blue asymmetrical behavior.
This effect is of increasing importance with  He and H$_2$  density,
and as a result, the full treatment of the infrared K doublet reveals strong effects on its opacity outside the
impact approximation.
For  a He and   H$_2$ perturber density larger than  $10^{19}$~cm$^{-3}$
(0.4 bar) the impact approximation breaks down and  is not applicable
to calculate the shape of the line profiles and their line core parameters.
}


\authorrunning{N.~F.~Allard \& J.~F.~Kielkopf}

\titlerunning{Asymmetrical shape of  the Near-Infrared K 4p-5s
  doublet perturbed by He and H$_2$}

  \maketitle
  \nolinenumbers
  
\section{Introduction}
The opacity of alkali atoms, most importantly of Na and K, plays a
crucial role in the atmospheres of brown dwarfs and exoplanets.
The wings of the sodium and potassium resonance lines
in the optical are particularly important because they serve as a source of
pseudo-continuum opacity, reaching into the near-infrared (NIR) wavelengths in
the case of potassium. The relevance of these alkali species for the
atmospheres of such self-luminous objects was studied and discussed in
detail in \citet{burrows2001}.
Studies of observed L- and T-dwarf spectra by \citet{liebert2000} and
\citet{burrows2001}  showed clearly  the importance of extended
K resonance line wings (766.5 and 769.9 nm) and pointed out the need for
spectral broadening calculations more accurate than Lorentzian profiles.
Understanding the shape of these lines is essential 
to modeling the transport of radiation from the interior.
Compared to the commonly used van der Waals broadening in the
impact approximation, major first improvements in the theoretical
description of pressure broadening
have been made by \citet{burrows2003} and \citet{allard2003}.
The K--He/H$_2$  line profiles have been investigated for the
 resonance lines in  \citet{allard2016b,allard2024}.
This paper is a continuation of  \citet{allard2025} where 
we determined the broadening of Na/K  by H$_2$ in
the  unified theory  at  H$_2$ densities larger
than $10^{21}$~cm$^{-3}$ and compare to the corresponding Lorentzian profiles.
We demonstrated that the simple treatment such as Lorentzian profiles
coming from impact  broadening theory  is only valid in the core of
the line not farther than a few half-widths as long as there is no overlap
between the core of the line and  quasi-molecular features in the wings due
to close collisions. 

In the present work we do a similar study of the K doublet at
1.243/1.252~$\mu$m in the J-band. Typically, near-infrared spectroscopic
observations of T dwarfs are characterized  by the presence of this
  doublet \citep{cushing2005, cushing2008}. 
In theoretical spectra  of T dwarfs, this K  doublet is seen for values
 of T$_{eff}$ ~700 K and above and  can  serve as useful indicators of
 the surface gravity.
 Medium resolution near-infrared spectral
 data for components of the Luhman 16A and B brown dwarf
 binary obtained by \citet{faherty2014,lodieu2015} show strong K~I
 absorption at 1.243, and 1.254~$\mu$m in both components.
The asymmetrical shape of this doublet in Luhman 16AB has been obtained
in high resolution by \citet{deregt2025}, however this asymmetry could be
identified as well in  spectra of Luhman 16AB
(private communication) obtained
 with the X-shooter cross-dispersed echelle spectrograph of 
 \citet{lodieu2015}.

The main purpose of this work is to analyze the  K line profiles of the
doublet by examining the development of the blue shoulder of the line
profile due to close  quasi-molecular unresolved line features 
and the consequent effects on the characteristics  of the line profiles.
In our work, a unified line shape theory and a set of atomic interaction
potential energies are used to model the entire line profile from the
impact-broadened line center to the far wing.
Our theoretical approach is based on the quantum theory of
spectral line shapes of~\citet{baranger1958a,baranger1958b} 
in an  adiabatic representation to include the degeneracy of atomic
levels \citep{royer1974,royer1980,allard1994}.
Complete details and the derivation of the theory 
are given in~\citet{allard1999}, a rapid account of the theory is given
in a recent paper \citep{allard2023}.
The molecular-structure calculations performed by
\citet{pascale1983} for  K--He are used for the adiabatic potential
of  the 5$s$ state and we use  a combination of
K--He ab initio  potentials of 
\citet{santra2005} at  short internuclear distance and those of
\citet{nakayama2001b} elsewhere  for the $4p$ state (see \citet{allard2024}).
Molecular data calculations  for the $4p$ and $5s$ states of  K--H$_2$ were
described in \citet{allard2016b}.
 
In Section~\ref{sec:asym}  we present a comparative  study of the K--He
and K--H$_2$ line profiles for the  typical temperature and pressure
structure for these stars.
 A full study of the doublet  allows us to  check
 the conditions of validity of the Lorentzian approximation used
in \citet{deregt2025} to model  the  high resolution spectrum of the
Luhman 16AB observed with the CRyogenic high resolution InfraRedEchelle
Spectrograph (CRIRES$^+$).
The unified line shape theory of \citet{allard1999} incorporating more accurate
potentials than van der Waals approximations allowed improvement of the
observed blue shift of the K doublet at 1.243/1.252~$\mu$m of Luhman 16AB.
In  Section~\ref{sec:param} we discuss the major importance of perturber
density in the broadening, shift and asymmetry measured
from calculated  profiles within  a unified line shape semi-classical
theory.

\begin{figure}
 \centering
\resizebox{0.46\textwidth}{!}
{\includegraphics*{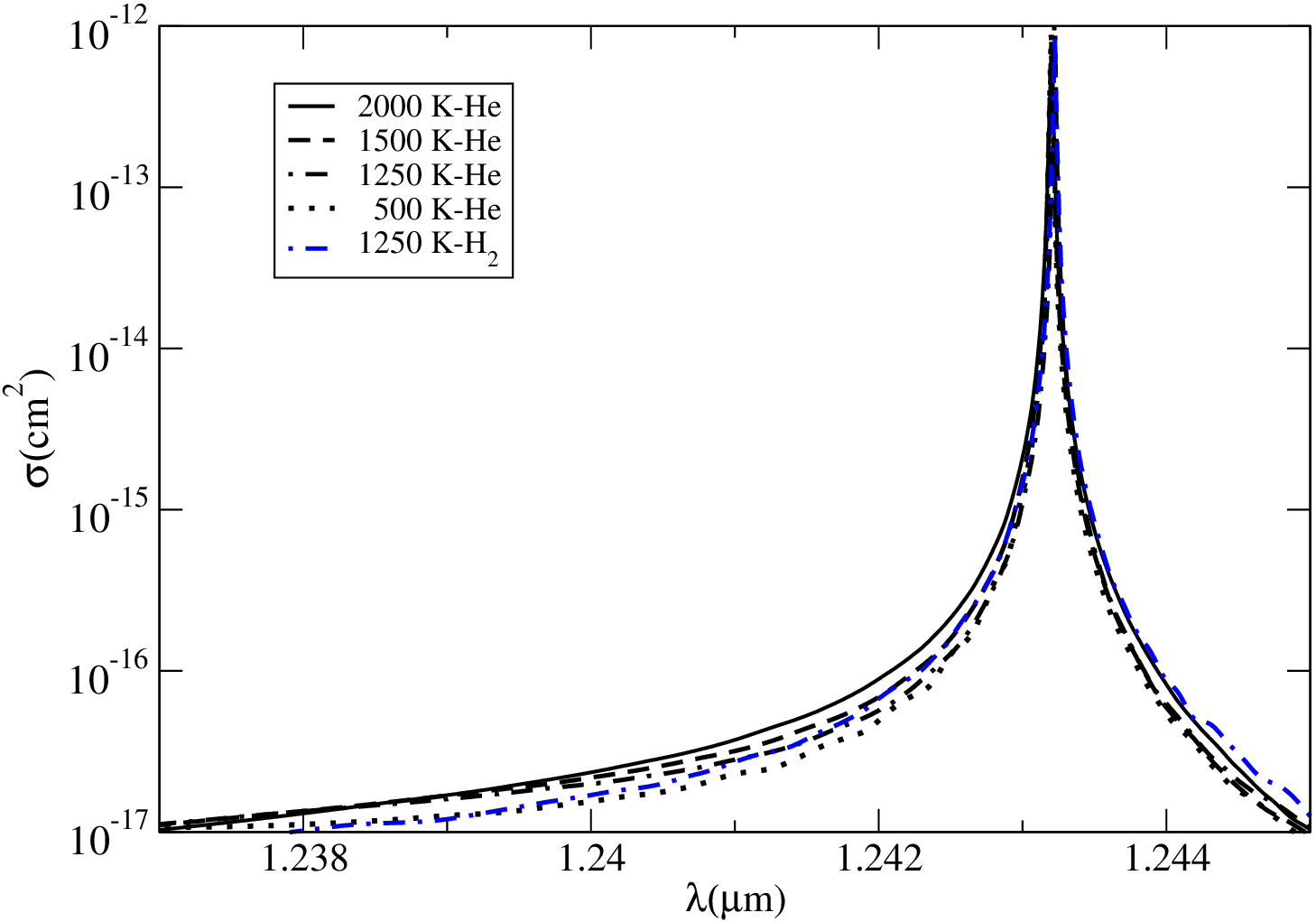}}
\caption  {Variation of the  absorption cross section, $\sigma$,
of  the  $4p\,^2P_{1/2}$--$5s\,^2S_{1/2}$ ($D1$)  component  of K perturbed by  He with temperature,)
(from top to the bottom \mbox{$T$  =2000, 1500, 1250 and 500$\ \mathrm{K}$}, 
n$_{\rm He}=10^{18}$~cm$^{-3}$). K--H$_2$ is plotted for
$T$=1250$\ \mathrm{K}$, n$_{\mathrm{H_2}}= 10^{18}$ ~cm$^{-3}$.}
\label{fig:varTD1}
\end{figure}

\begin{figure}
 \centering
\resizebox{0.46\textwidth}{!}
{\includegraphics*{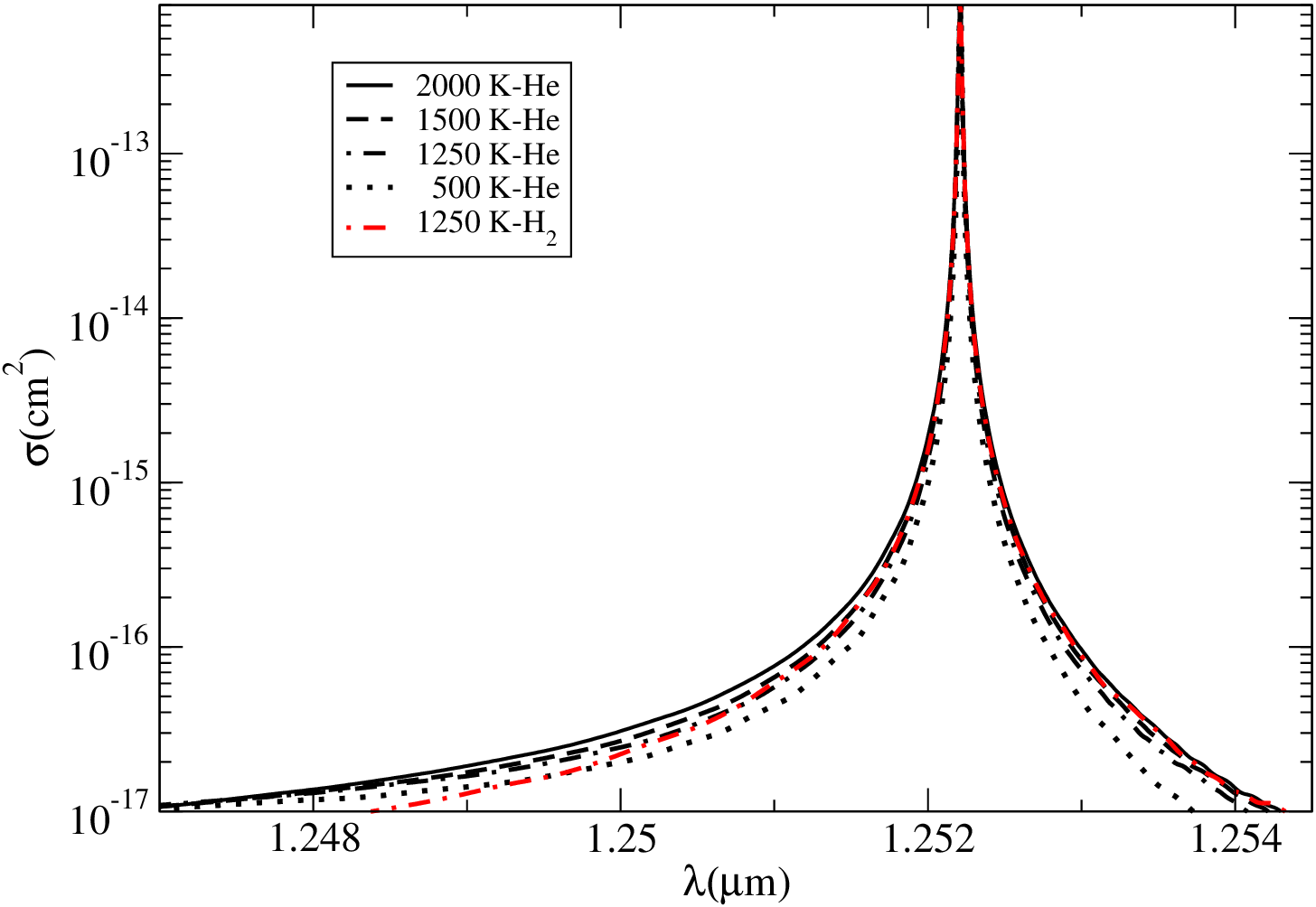}}
\caption  {Variation of the  absorption cross section, $\sigma$,
of  the  $4p\,^2P_{3/2}$--$5s\,^2S_{1/2}$  ($D2$) component  of K perturbed by  He with temperature,
(from top to the bottom \mbox{$T$  =2000, 1500, 1250 and 500$\ \mathrm{K}$}, 
n$_{\rm He}$=10$^{18}$~cm$^{-3}$). K--H$_2$ is plotted for
$T$=1250$\ \mathrm{K}$, n$_{\mathrm{H_2}} = 10^{18}$ ~cm$^{-3}$.}
\label{fig:varTD2}
\end{figure}

\section{Asymmetrical shape of the  K $4p$--$5s$ doublet}\label{sec:asym}
The main purpose of this section  is to analyze the  line profiles
of  the $4p\,^2P_{1/2}$--$5s\,^2S_{1/2}$ (1.243~$\mu$m designated $D1$) and $4p\,^2P_{3/2}$--$5s\,^2S_{1/2}$ (1.252~$\mu$m designated $D2$) doublet by examining the development of quasi-molecular
effects in  K--He/H$_2$ collisions. The transitions share a common $5s\,^2S_{1/2}$ upper state. They are distinguished by lower states that are the upper states of the K resonance lines. The designations we use in the following associate each infrared line with its corresponding resonance transition.   
Both line  profiles of the K doublet exhibit a blue asymmetrical behavior 
illustrated in Figs.~\ref{fig:varTD1}--\ref{fig:varTD2}
for a very low He density (n$_{\rm He}$=10$^{18}$~cm$^{-3}$).
These calculations span the range
$T_\mathrm{eff}$=500$\ \mathrm{K}$ to 2000~$\ \mathrm{K}$ and
take into account  the spin-orbit coupling as described
by  \citet{allard2006}.
K--H$_2$ line  profiles are overplotted for $T$=1250$\ \mathrm{K}$
for the same H$_2$
 perturber density, n$_{\mathrm{H_2}}$ = $10^{18}$ ~cm$^{-3}$.

\begin{figure}
  \centering
  \vspace{8mm}  
   \includegraphics[width=0.9\linewidth]{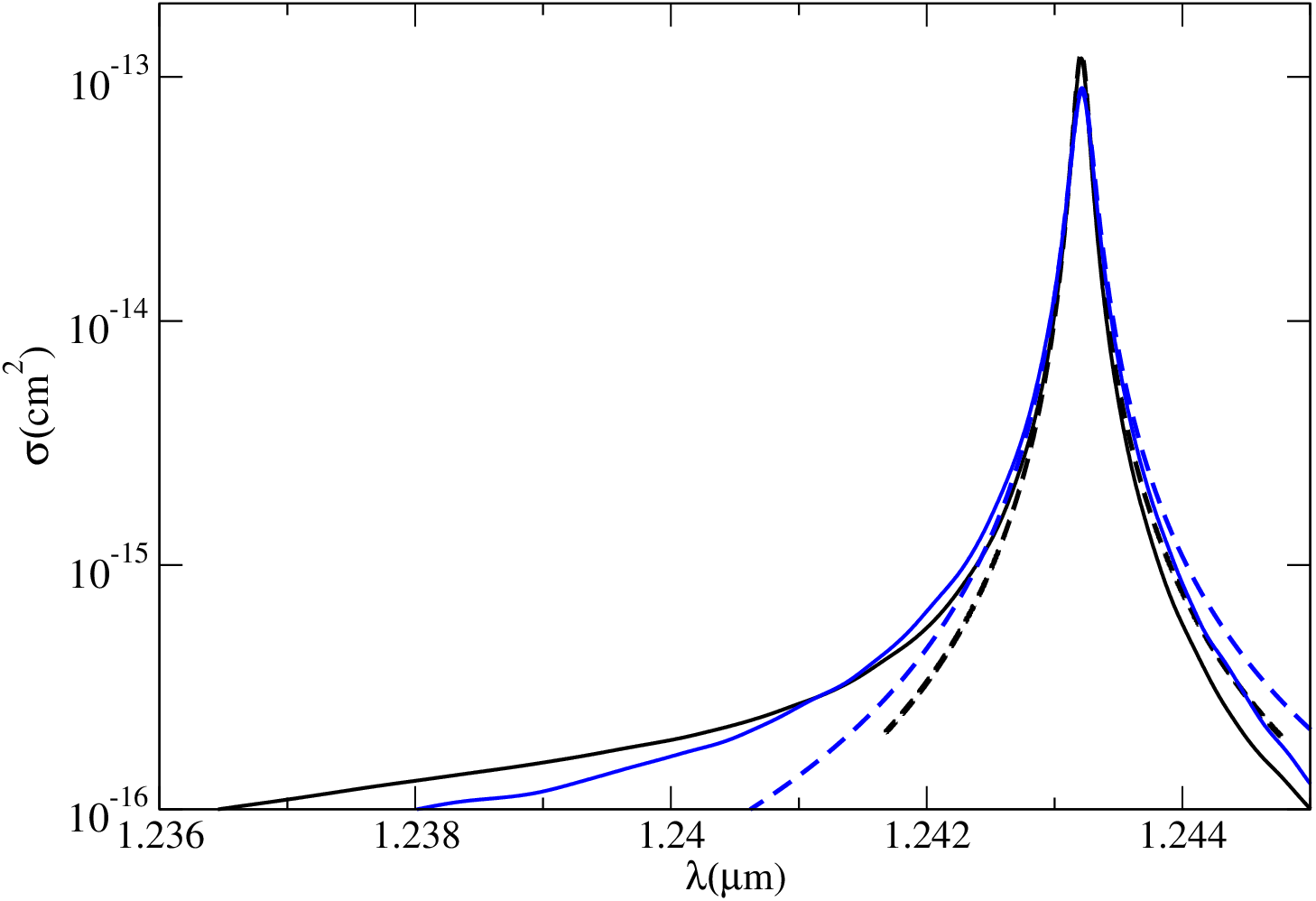}
   \caption{Comparison of the absorption cross section, $\sigma$,
     of  the $4p\,^2P_{1/2}$--$5s\,^S_{1/2}$ ($D1$) component  of K perturbed by  H$_2$  (blue  curve)
     with K perturbed by He collisions (black curve) 
     at $T = 1250~K$ and $n_{\mathrm{He}} = n_{\mathrm{H_2}} = 10^{19}$~cm$^{-3}$. 
The Lorentzian approximation is overplotted for comparison (dashed curves).}
   \label{fig:KH2HeD1}
\end{figure}

\begin{figure}
  \centering
  \vspace{8mm}  
   \includegraphics[width=0.9\linewidth]{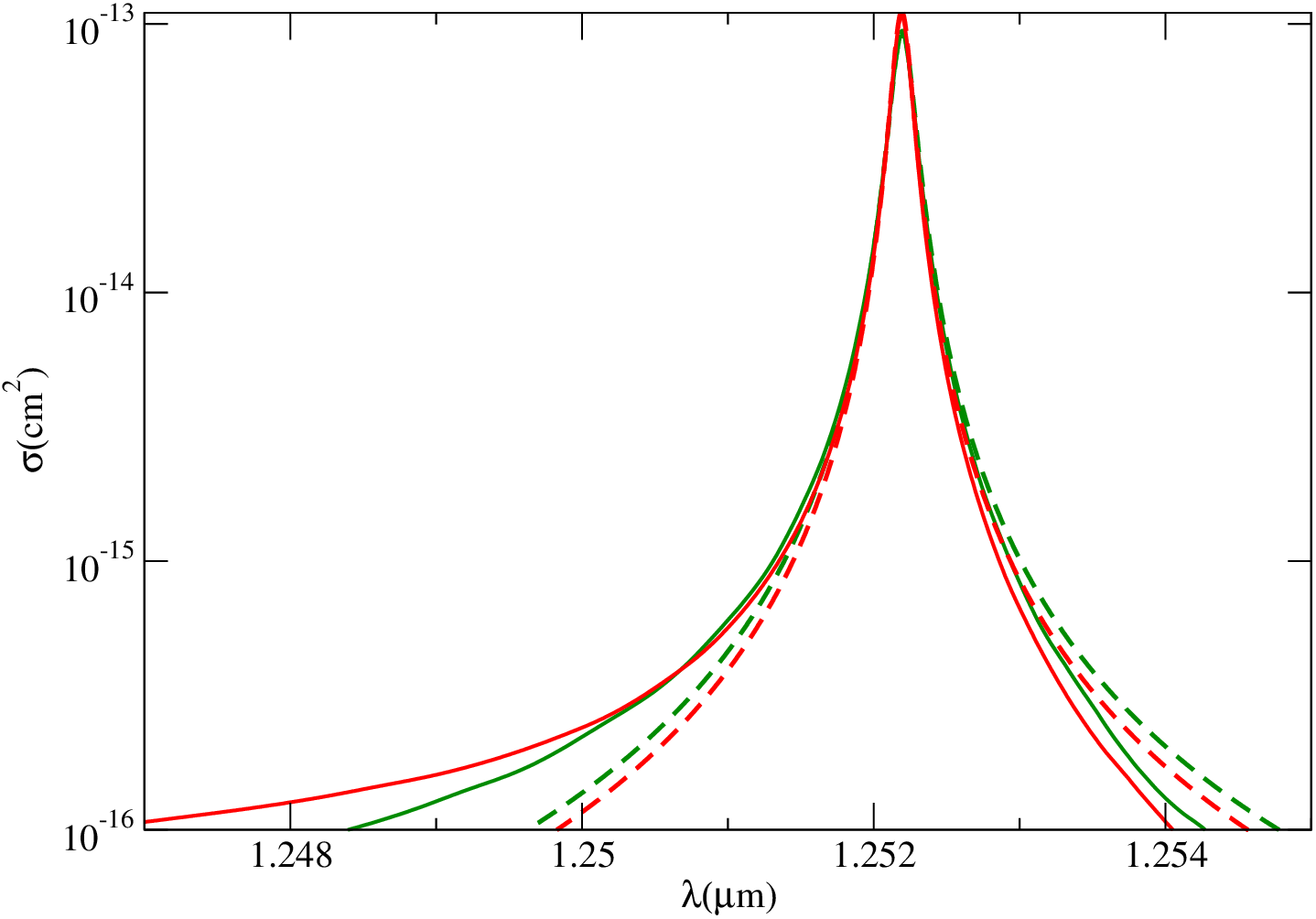}
   \caption{Comparison of the absorption cross section, $\sigma$,
     of  the  $4p\,^2P_{3/2}$--$5s\,^2S_{1/2}$ ($D2$)  component  of K perturbed by  H$_2$  (red  curve)
     with K perturbed by He collisions (green curve) 
     at $T = 1250~K$ and $n_{\mathrm{He}} = 10^{19}$~cm$^{-3}$. 
The Lorentzian approximation is overplotted for comparison (dashed curves)}
   \label{fig:KH2HeD2}
\end{figure}

\begin{figure}
  \centering
  \vspace{8mm}  
   \includegraphics[width=0.9\linewidth]{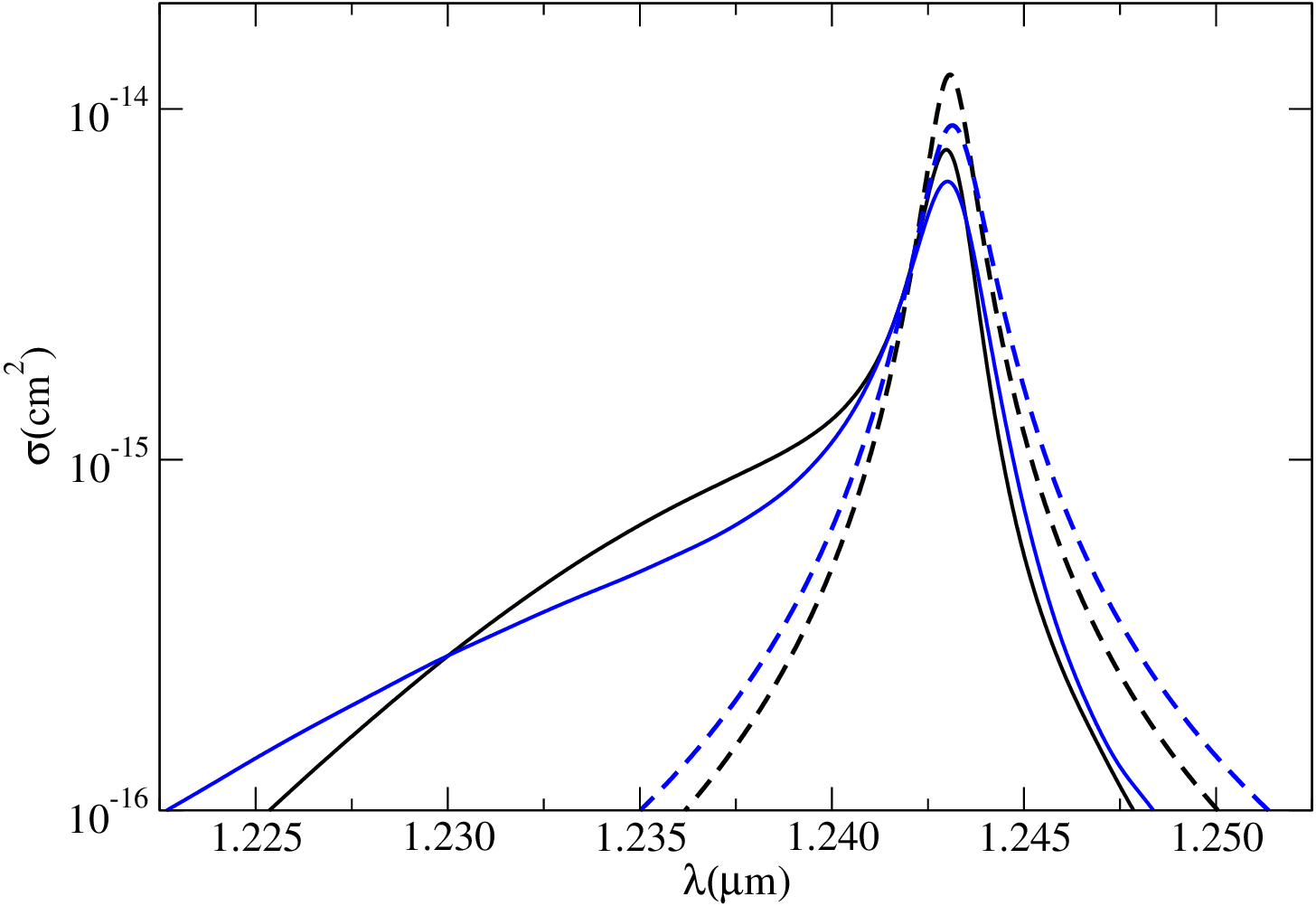}
   \caption{Comparison of the absorption cross section, $\sigma$,
     of  the $4p\,^2P_{1/2}$--$5s\,^2S_{1/2}$ ($D1$)   component  of K perturbed by  H$_2$  (blue  curve)
     with K perturbed by He collisions (black curve) 
     at $T = 1250~K$ and   $n_{\mathrm{He}} = n_{\mathrm{H_2}} = 10^{20}$~cm$^{-3}$. 
The Lorentzian approximation is overplotted for comparison (dashed curves)}
   \label{fig:KH2HeD1dens20}
\end{figure}

\begin{figure}
  \centering
  \vspace{8mm}  
   \includegraphics[width=0.9\linewidth]{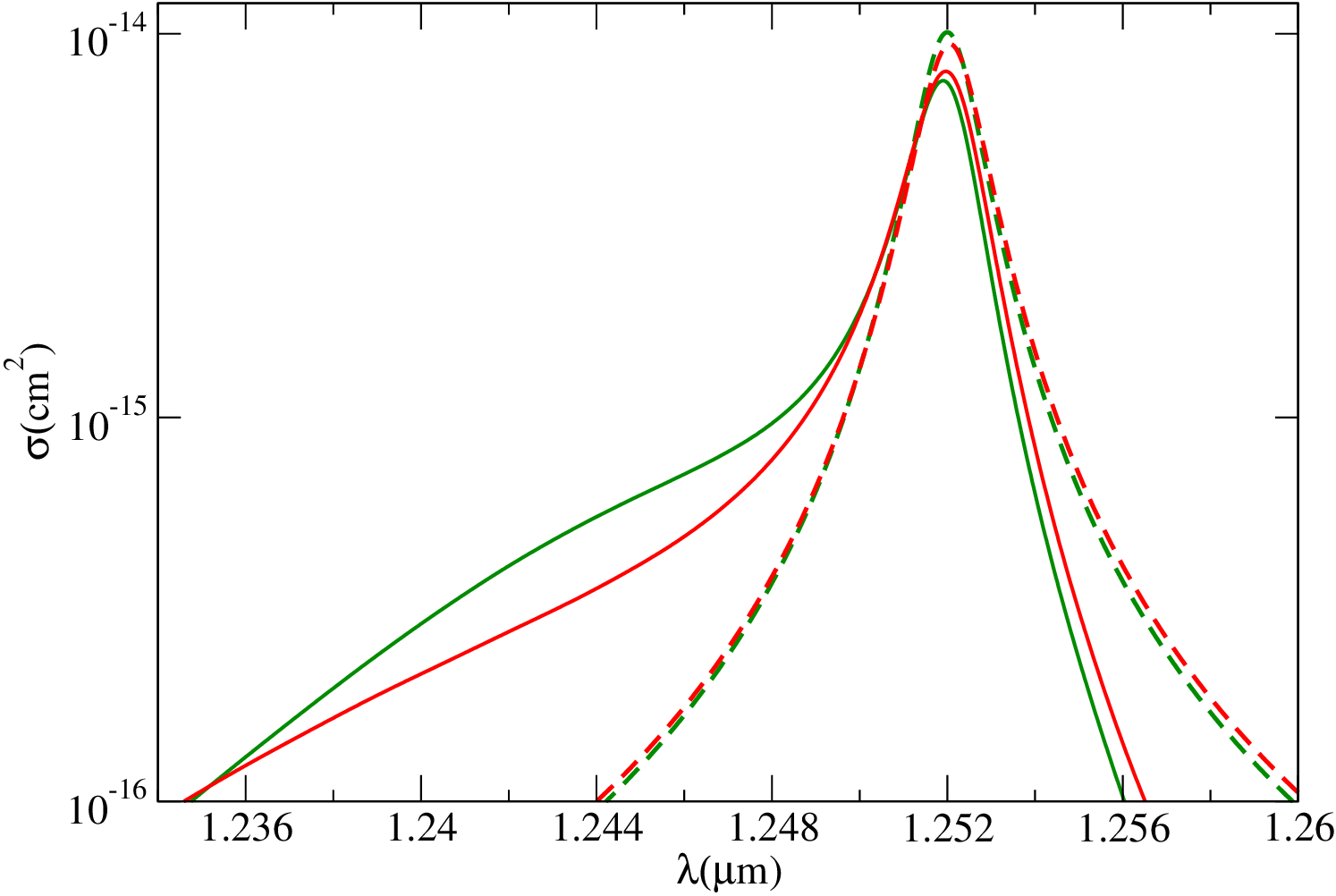}
   \caption{Comparison of the absorption cross section, $\sigma$,
     of  the  $4p\,^2P_{3/2}$--$5s\,^2S_{1/2}$ ($D2$)  component  of K perturbed by  H$_2$  (red  curve)
     with K perturbed by He collisions (green curve) 
     at $T = 1250~K$ and   $n_{\mathrm{He}} = n_{\mathrm{H_2}} = 10^{20}$~cm$^{-3}$. 
The Lorentzian approximation is overplotted for comparison (dashed curves)}
   \label{fig:KH2HeD2dens20}
\end{figure}

Figures~\ref{fig:varTD1} and \ref{fig:varTD2} reveal that increasing temperature at a fixed perturber density increases opacity in the line wings.  Figures~\ref{fig:KH2HeD1} to \ref{fig:KH2HeD2dens20} show a  far more significant dependence on density at a fixed temperature. 
The K--He/H$_2$ collisional line profiles shown in
Figs.~\ref{fig:KH2HeD1}--\ref{fig:KH2HeD2dens20}
are evaluated for
$n_{\mathrm{He}}= n_{\mathrm{H}_2} = 10^{19}$ and $10^{20}$~cm$^{-3}$,
$T_\mathrm{eff}=1250$~$\mathrm{K}$. They illustrate the  similarity of the line shapes of the doublet components, and that
the H$_2$ blue line wings have approximately the same peak strength as 
the ones induced by He. In the following we will restrict the study 
to $T_\mathrm{eff}=1250$~$\mathrm{K}$ to explore the dependence on density.

\begin{figure}
  \centering
  \vspace{8mm}  
\resizebox{0.46\textwidth}{!}{\includegraphics*{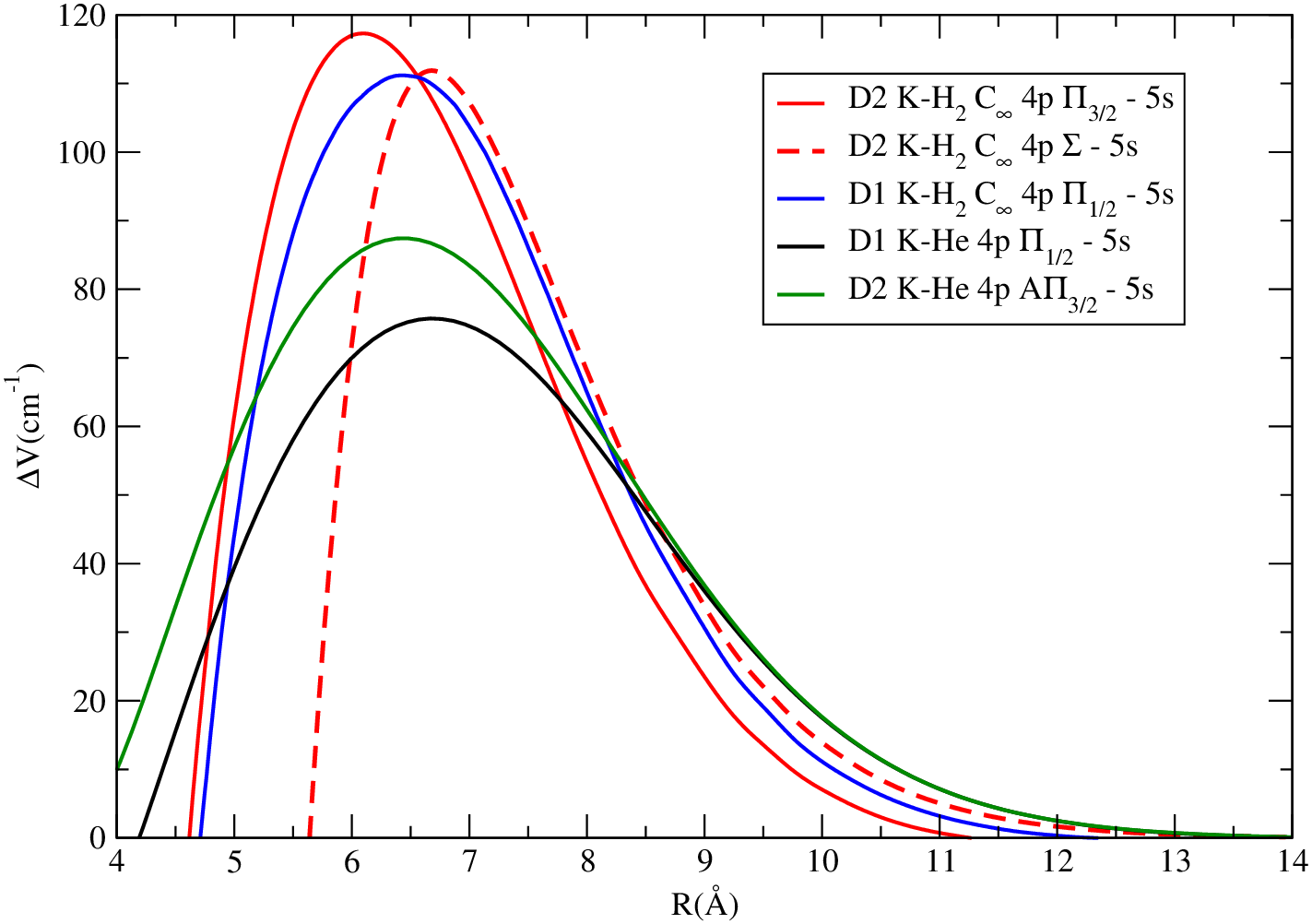}}
\caption  {Potential differences $\Delta V(R)$  corresponding to the 
  $4p$--$5s$ transitions:  K--He (black curve for
  $D1$, green curve for $D2$) and K--H$_2$ (blue curve for $D1$,
  red solid for the $4p\,\Pi$  and red dashed for the $4p\,\Sigma$ lower states determining $D2$).
\label{fig:diffpotKHeKH2}}
\end{figure}

\begin{figure}
  \centering
  \vspace{8mm}  
   \includegraphics[width=0.9\linewidth]{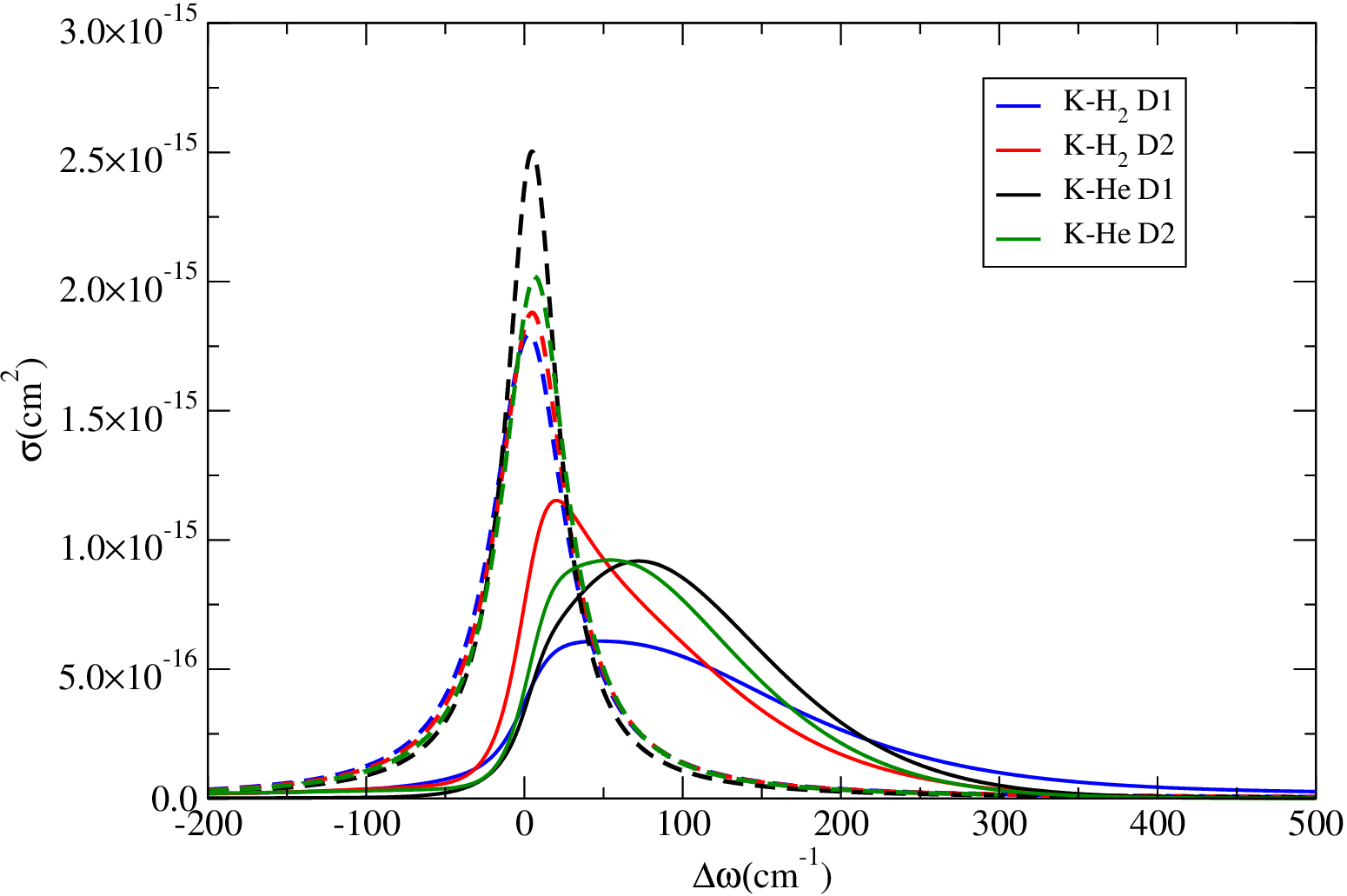}
   \caption{Comparison of the absorption cross section, $\sigma$,
     of  the $D1$ (blue  curve)  and  $D2$ components  of K perturbed by  H$_2$
     (red  curve)
     with the $D1$ (black  curve)  and  $D2$  (green curve) components of 
     K perturbed by He collisions 
     at $T = 1250~K$ and
     $n_{\mathrm{He}}$ = $n_{\mathrm{H_2}} = 5\times 10^{20}$~cm$^{-3}$. 
     The abscissa in the plot is given in wavenumbers (cm$^{-1}$), which reverses ``blue'' and ``red'' while emphasizing the effect of the potential differences on the spectrum.   
   $\Delta\omega$ is  relative to  the unperturbed atomic line
     in cm$^{-1}$, and comparable to the energy difference $\Delta V$ of
   Fig. \ref{fig:diffpotKHeKH2} in the same units. The Lorentzian approximation is overplotted for comparison (dashed curves).}
   \label{fig:KH2HeD1dens5e20}
\end{figure}

In radiative collision transitions it is the difference between the
final and initial state potential energies that determines the frequency
and the energy of a single photon emitted or absorbed by the interacting atoms.
The interpretation of the asymmetrical shape of the 
 \mbox {$4p \rightarrow 5s$} doublet
 requires us to study the potential energy difference
 $\Delta V$ (Fig.~\ref{fig:diffpotKHeKH2}).
 The unified theory \citep{anderson1952,allard1978} predicts that there will
be line satellites centered periodically at 
frequencies corresponding to the extrema of the difference potential
between the upper and lower states.
 The difference potential maxima  vary between 70 and
 120~cm$^{-1}$.  These low  maxima are responsible for  unresolved
line satellites resulting in the asymmetrical profiles. 
In comparison,  $\Delta V_{\rm max}$ for the K resonance lines
perturbed by He and H$_2$ were 1300 and  1500~cm$^{-1}$ respectively, 
which lead to well-resolved line satellites \citep{allard2024,allard2016b}.
At densities well below  $10^{19}$~cm$^{-3}$ typical line profiles in the
core of the line are Lorentzian.
With increasing perturber densities the blue tails
(Figs.~\ref{fig:KH2HeD1}--\ref{fig:KH2HeD2}) become a shoulder which
climbs up along the blue side of the main line and distorts the line
shape (Figs.~\ref{fig:KH2HeD1dens20}-\ref{fig:KH2HeD2dens20}).
The unresolved line satellites   lead to a blue displacement of the core
of the line more important than those due to impact shifts (see 
  Fig.~\ref{fig:shiftlowdens} in the next section).
There is a linear variation of the strength of
the blue wing with perturber  density as long as
 \mbox{ n$_{\rm He}$, n$_{\rm H2}$ $\leq$ 10$^{20}$  cm$^{-3}$}.
There is a dramatic change when the density gets larger 
than $10^{20}$~cm$^{-3}$: the core  becomes markedly non-Lorentzian and
asymmetrical in Fig.~\ref{fig:KH2HeD1dens5e20}.

\section{Study of the line parameters}
\label{sec:param}
Since  the resulting 
line profile in a model atmosphere calculation is the integration of 
the flux in all layers from the deepest
to the uppermost, it is important that the centers be adequately 
represented; that is, they can be non-Lorentzian at  high densities 
of the innermost layers, 
while Lorentzian in the upper atmosphere. However, they are characterized
   with different widths 
than predicted by the
hydrogenic van der Waals approximation, which is usually 
used for the cores, as was 
emphasized by \citet{allard2007c} and \citet{peach2011}.
The core shape does not depend on the assumption that the
interaction is given by a van der Waals approximation, but  the
width and shift of the core are very dependent on the nature of
the interactions between the atoms.  
At sufficiently low densities of helium or molecular hydrogen, in the upper
atmospheres,  the symmetrical center of the spectral line  can be defined
 by the two line impact parameters:  
 the width $w_{\rm imp}$, and the shift $d_{\rm imp}$ of the main line.
They are reported in Eqs.~12 and 13 of \citet{deregt2025}.

\begin{figure}
 \centering
\resizebox{0.46\textwidth}{!}
{\includegraphics*{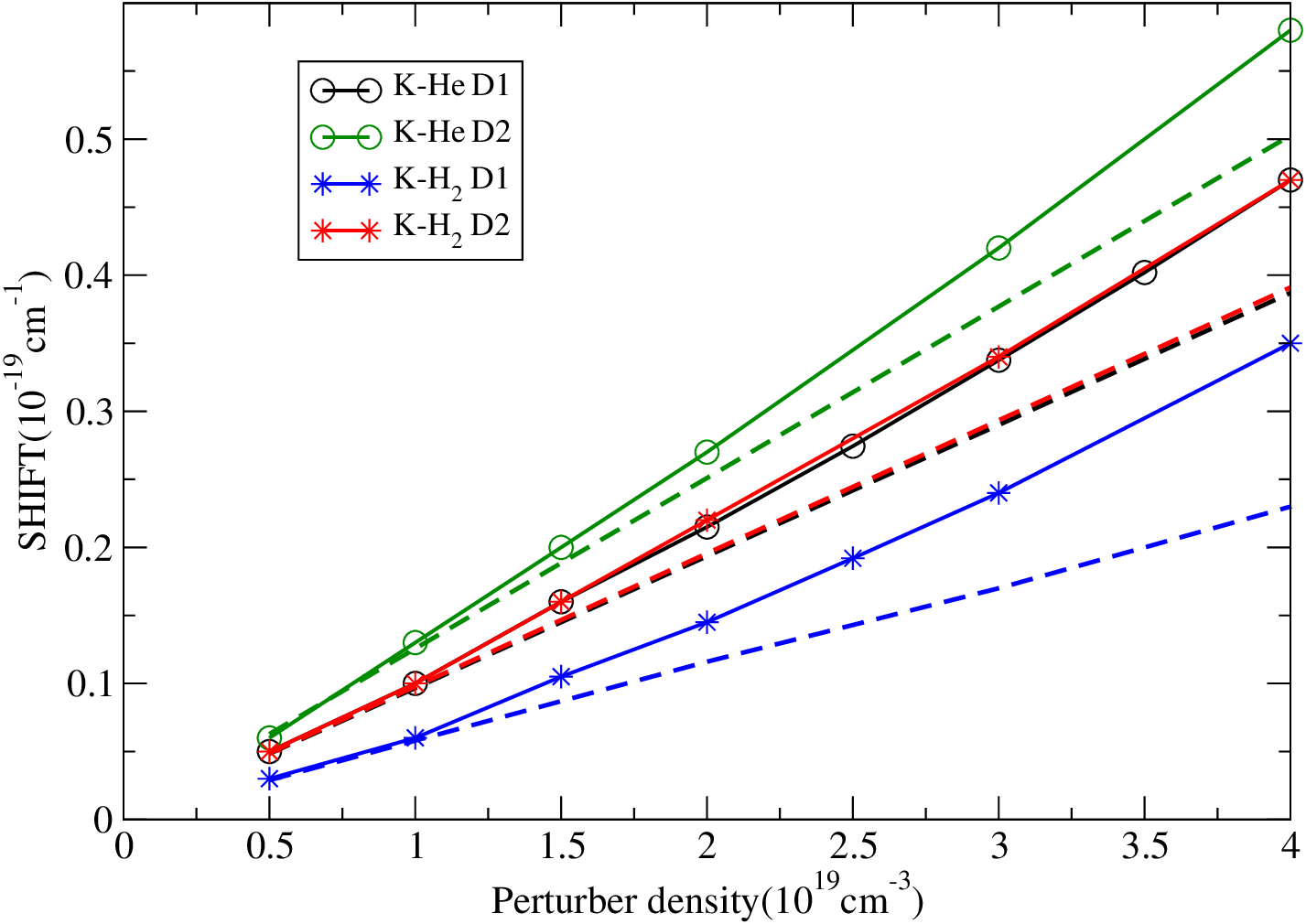}}
\caption  {The  shift of the K--He  $D1$(black), $D2$ (green),
 the KH$_2$  $D1$ (blue) and  $D2$ (red) 
  spectral lines as a function of He and H$_2$ density.
  The impact approximation  (dashed line)  shown for comparison
  is invalid for all but the lowest densities.}
\label{fig:shiftlowdens}
\end{figure}

\begin{figure}
 \centering
\resizebox{0.46\textwidth}{!}
{\includegraphics*{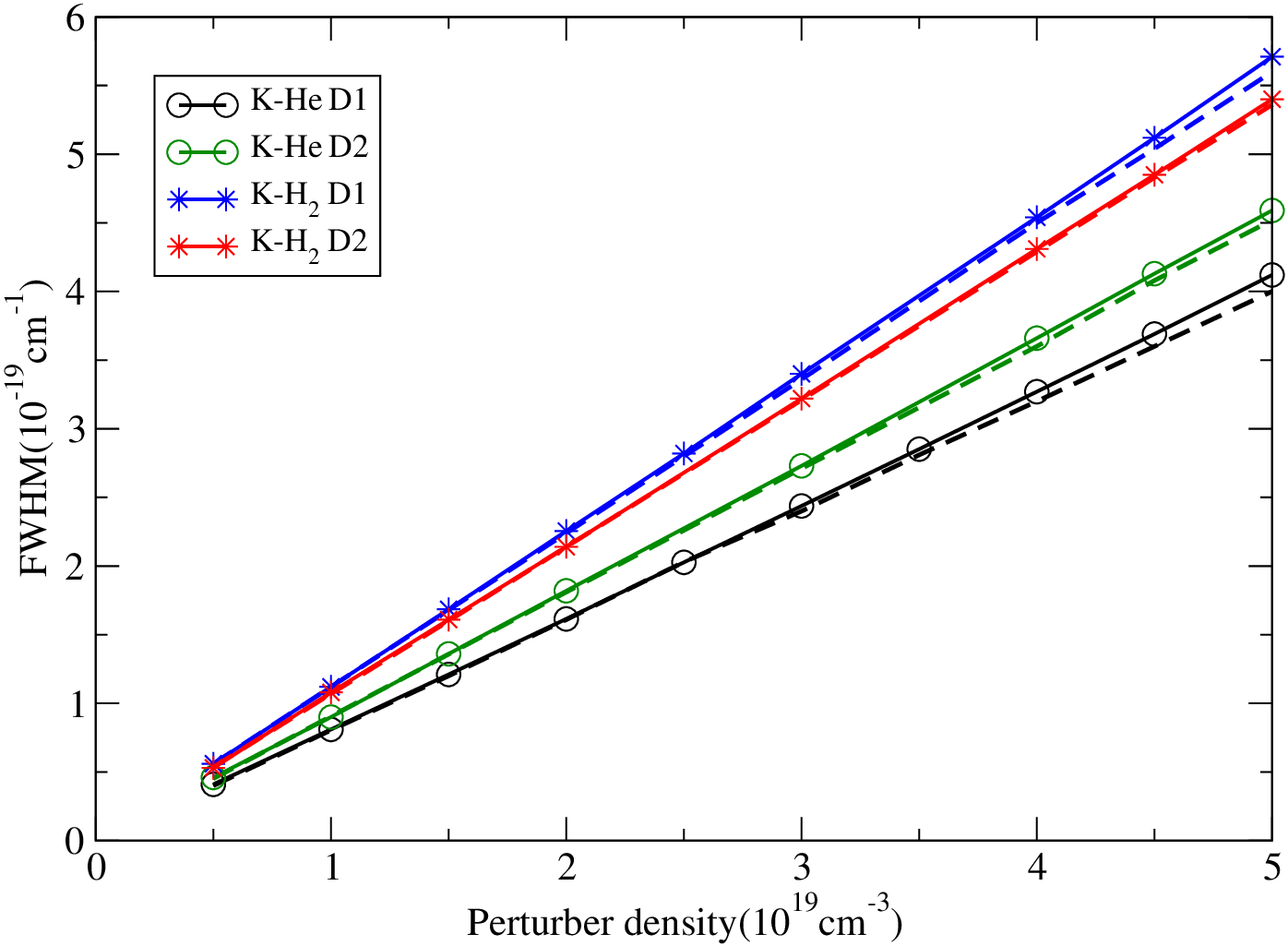}}
\caption  {FWHM  of the K--He  $D1$(black), $D2$ (green),
 the K--H$_2$  $D1$ (blue) and  $D2$ (red) 
  spectral lines as a function of He and H$_2$ density.
  The impact approximation  (dashed line)  shown for comparison
  is invalid for all but the lowest densities.}
\label{fig:widthlowdens}
\end{figure}

\begin{figure}
 \centering
\resizebox{0.46\textwidth}{!}
{\includegraphics*{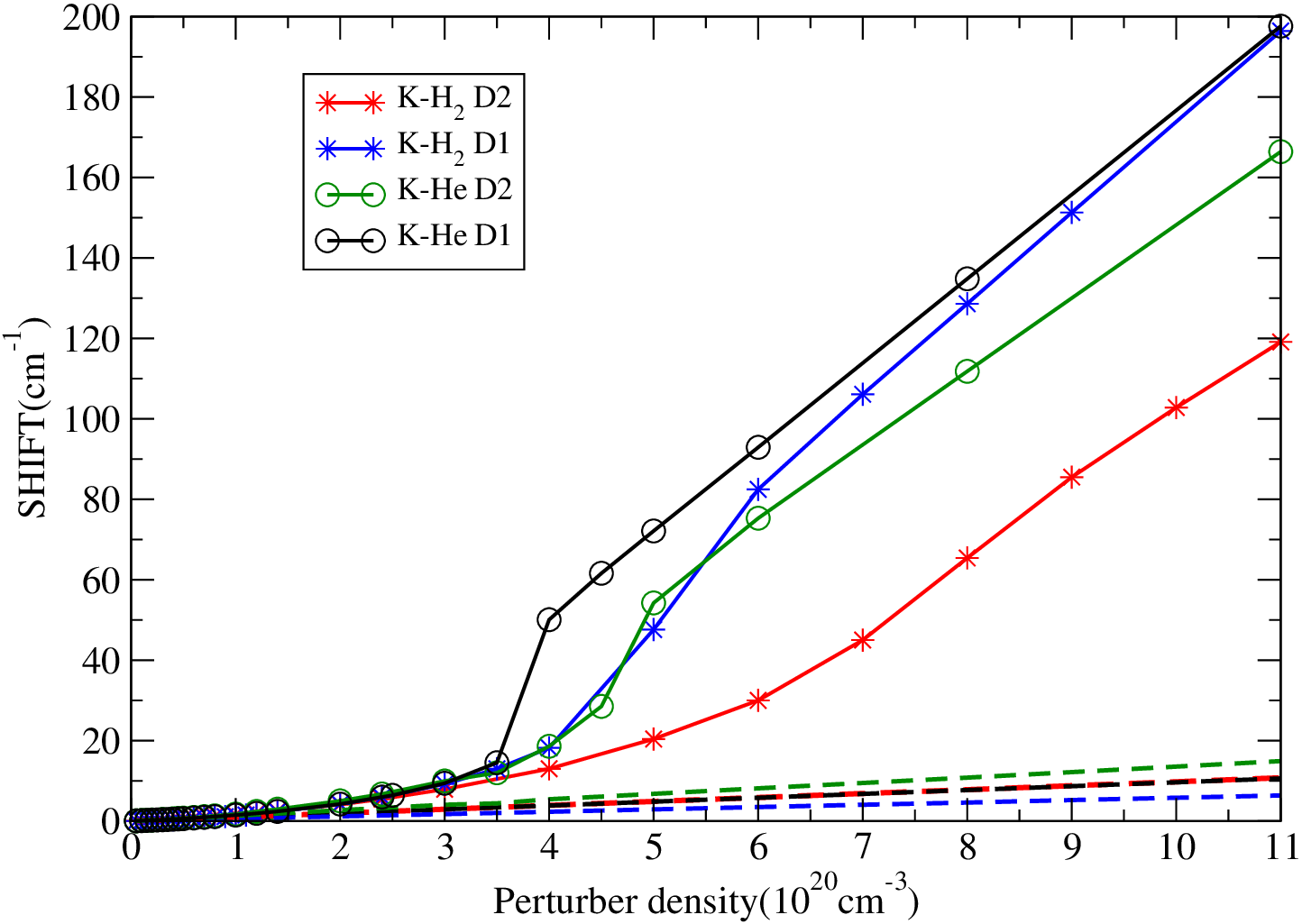}}
\caption  {The  shift of the K--He  $D1$(black), $D2$ (green),
 the K--H$_2$  $D1$ (blue) and  $D2$ (red) 
  spectral lines as a function of He and H$_2$ density.
  The impact approximation  (dashed line)  shown for comparison
  is invalid for all but the lowest densities.}
\label{fig:shift}
\end{figure}

\begin{figure}
 \centering
\resizebox{0.46\textwidth}{!}
{\includegraphics*{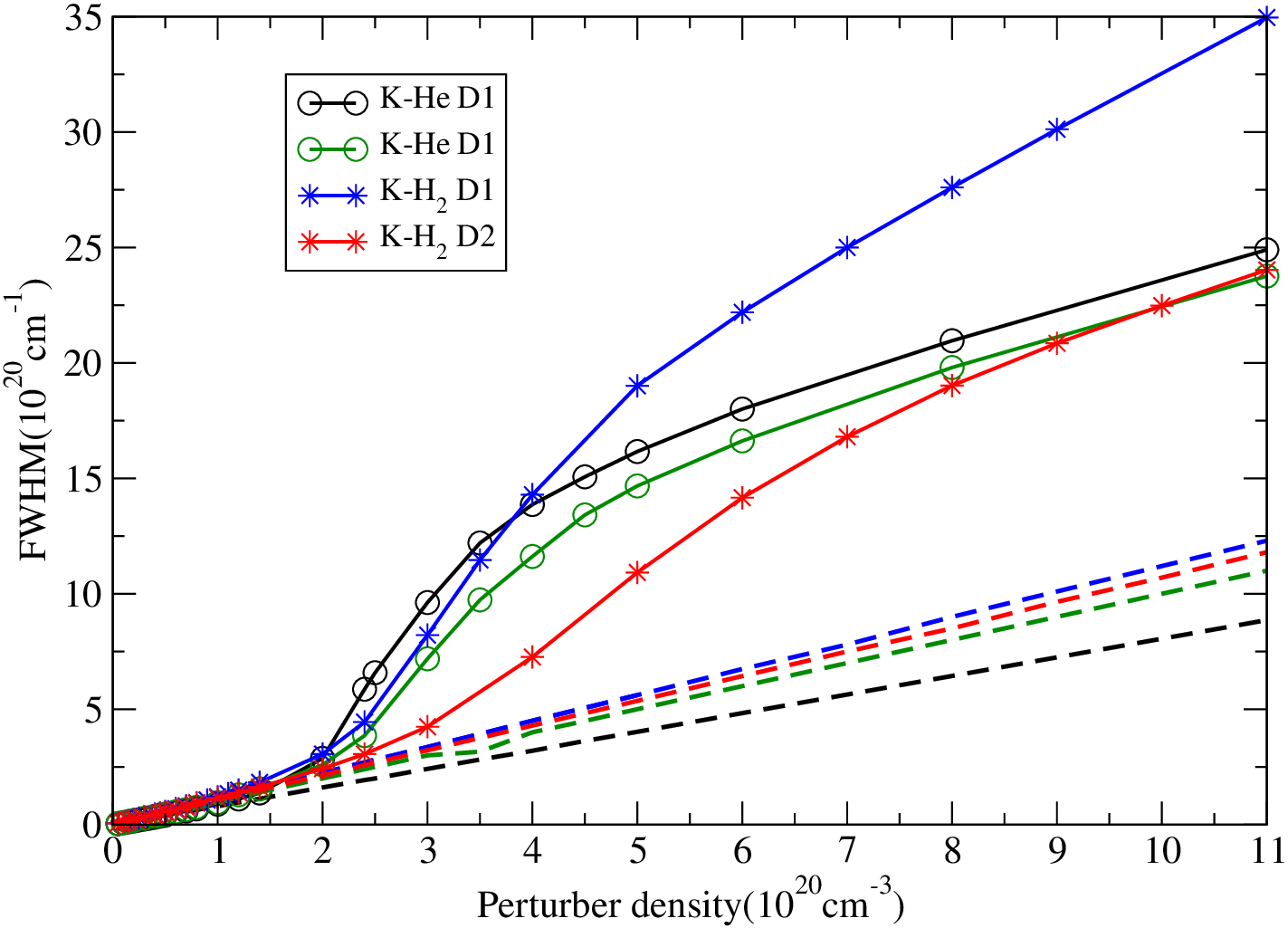}}
\caption  {FWHM  of the K--He  $D1$(black), $D2$ (green),
 the K--H$_2$  $D1$ (blue) and  $D2$ (red) 
  spectral lines as a function of He and H$_2$ density.
  The impact approximation  (dashed line)  shown for comparison
  is invalid for all but the lowest densities.}
\label{fig:width}
\end{figure}

\begin{figure}
 \centering
\resizebox{0.46\textwidth}{!}
{\includegraphics*{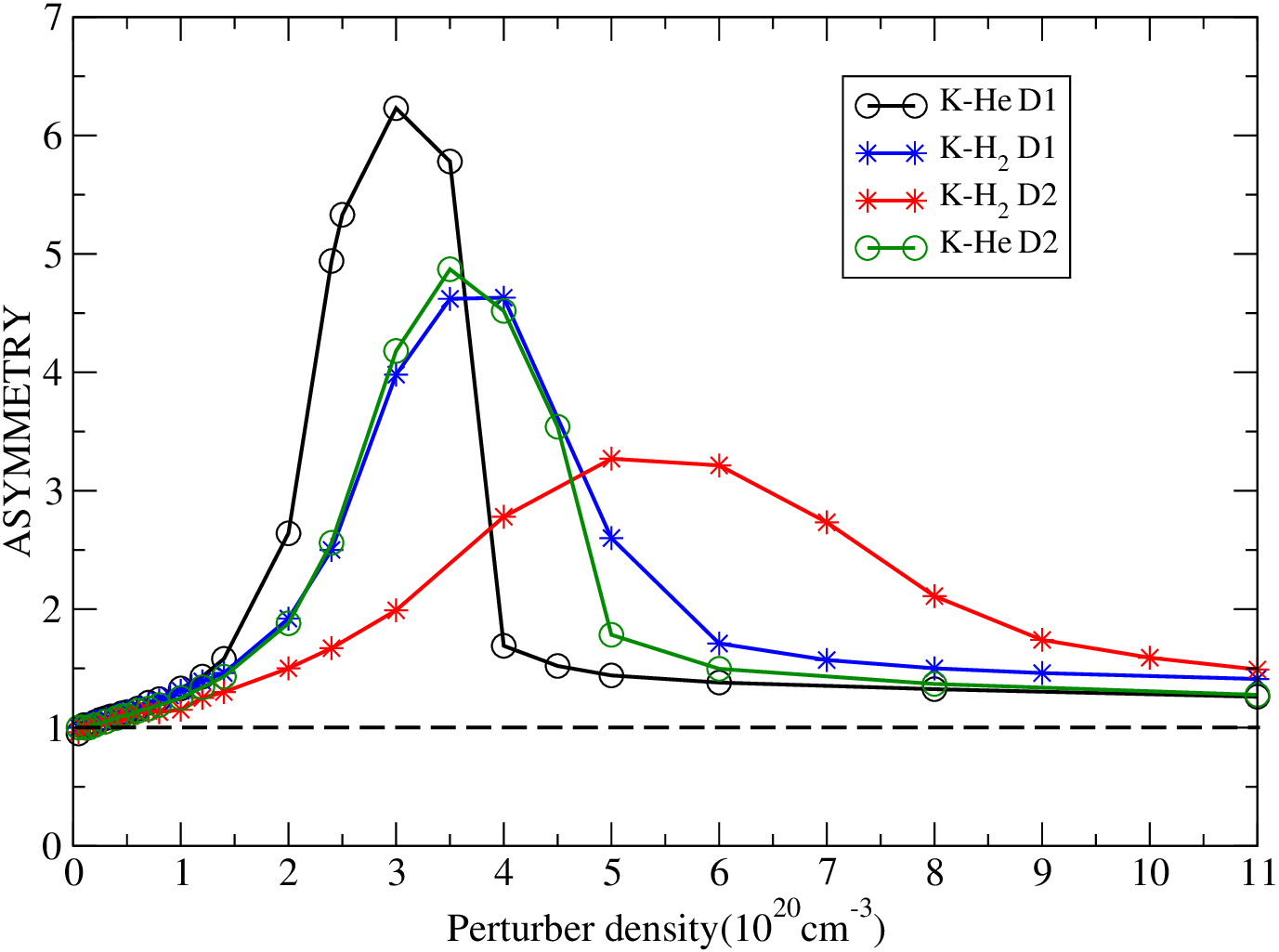}}
\caption  {Dimensionless asymmetry  of the K--He  $D1$(black), $D2$ (green),
 the K--H$_2$  $D1$ (blue) and  $D2$ (red) 
  spectral lines as a function of He and H$_2$ density.}
\label{fig:asym}
\end{figure}

 Clearly understanding the variation of shape 
 of the $4p$--$5s$ line with increasing perturber  density 
allows us to understand the possible important  blue shift of the doublet.
The simplest way of characterizing a line shape is in terms of the
parameters, $w$, the full width at half maximum (FWHM), $d$, the shift
and the asymmetry (which is computed by taking the ratio of the
width at half of the maximum intensity of the blue side to that
of the red side).  The dependence of the line parameters on density
has two origins.
First in the limit of low densities, the main line
is very sharp, the blue shoulder  is small and well-separated from the
core of the line (Figs.~\ref{fig:varTD1}-\ref{fig:varTD2}).
 The shift  changes linearly with density at
 the same rate as the impact value (Fig.~\ref{fig:shiftlowdens}).
 The same is true for the width  (Fig.~\ref{fig:widthlowdens}).
 The asymmetry parameter is almost one as expected in the impact
 approximation (Fig.~\ref{fig:asym}).
 The limit of validity of the impact theory occurs at ~10$^{19}$~cm$^{-3}$. 
Second the amplitude of the main line decreases, whereas the  
 shoulder climbs up along the blue side of the line to
become more important than the main line. When measured to the maximum of
the profile the parameters are now measured from the blue shoulder.
Hence, a ``discontinuity'' appears on the shift and the width
(Figs.~\ref{fig:shift}-\ref{fig:width}). 
 An analogous ``discontinuity'' appears on the asymmetry curve
 (Fig.~\ref{fig:asym}).
 The reason for this transition to very non-linear behavior of
 the line parameters is that
 we are first measuring the width proper at lower density. Then, 
as the density increases and the blue shoulder  grows
to more than one-half the height of the line, 
we are measuring the combined width of line and unresolved line satellites.
  Figure~\ref{fig:KH2HeD1dens5e20} shows that the line profiles of the  lines have orders of magnitude more absorption in the blue wing  than
  a Lorentzian when
  $n_{\mathrm{He}}= n_{\mathrm{H2}}$ = 5$\times$~10$^{20}$~cm$^{-3}$.
  In this illustration, $\sigma$ is a function of 
  $\Delta \omega$. The frequency  difference (here in cm$^{-1}$) is relative to the
  unperturbed spectral line center and the blue wing is to the right.
The disagreement with a  Lorentzian profile depends on the value of
$\Delta V_{\mathrm{max}}$ in Fig.~\ref{fig:diffpotKHeKH2} responsible of the
appearance of the quasi-molecular features.
  In comparison, for the Na/K resonance lines, the asymmetrical shape arises
when the H$_2$ density reaches 10$^{22}$~cm$^{-3}$ \citep{allard2025}.
The impact approximation breaks down at a
density limit that is not necessarily very high.
The width, shift, and asymmetry
of the lines are then very dependent on the nature of
the interactions between the atoms at short and intermediate distance.
We conclude that when an additional asymmetry is observed the impact approximation does not apply and cannot be used to calculate the line core.
The asymmetry signals the presence of a line
satellite within the core of
the line profile.  For excited states these effects arise even at low density, 
as noted in Fig.~6 of \citet{allard2012b}  for the
$3p$--$4s$ transitions of  Na perturbed by H$_2$.
Table~1 of the impact line parameters of Na--H$_2$  for the $3p-4s$
transition  of \citet{allard2012b}  and  of K--H$_2$ for the 
$4p-5s$ transition in \citet{deregt2025} are valid only at densities below
$10^{19}$~cm$^{-3}$ (0.3 bar).

\section{Conclusion}

 A similar analysis presented in \citet{allard2025}
 for the resonance lines of Na ($3s$--$3p$) and K ($4s$--$4p$) perturbed by H$_2$ is done
 for the $4p$-$5s$ doublet. 
  The full study of the doublet allows us to  check  the conditions of
validity of the Lorentzian approximation used
in \citet{deregt2025} to model  the  spectrum of Luhman 16AB.
We have found that the 
\mbox{$4p_{1/2} \rightarrow  5s$} and \mbox{$4p_{3/2} \rightarrow  5s$}
lines  are  totally asymmetrical outside the line core which
stops to be a Lorentzian profile when  the He and   H$_2$ perturber density
become larger than  $10^{19}$~cm$^{-3}$ (0.4 bar) because of 
the presence of unresolved features in the core of the line profile.
The accurate computation of the  K $4p$--$5s$ doublet for  perturber density
larger than  $10^{19}$~cm$^{-3}$ requires use of a Fourier transform
of the autocorrelation function inside the model atmosphere code.
A striking example is given Fig.~13 of \citet{allard2016a} for the
triplet $3p$-$4s$ Mg lines in cool DZ white dwarfs.
 It clearly illustrates how the non-Lorentzian 
shapes can be outside the limited range of validity of the impact 
approximation usually used in stellar and planetary atmosphere modeling.
Previously, \citet{allard2013e}  discussed the major importance 
of blue satellites associated with He-He collisions that are close and blended with the line core. 
The line profiles of the  $2p - 3s$ transitions
exhibit a similar blue asymmetrical behavior.
The deepest photospheric layers in \citet{deregt2025}
achieve  H$_2$  density of n$_{\mathrm{H_2}}$ $\sim3\times10^{19}$~cm$^{-3}$,
this is certainly not large enough to account for the asymmetrical shape of
the observed spectra of Luhman 16AB. Only a correct treatment of inclusion of
the  K~I $4p$--$5s$ line profiles in the atmosphere models could improve the
determination of the surface gravity.

Analogous laboratory measurements of collision broadening by xenon of excited $s$
states of thallium has been studied with a high resolution spectrometer by
\cite{kielkopf1983}.
The spectra had excellent signal-to-noise ratio, and were analysed
directly with profiles calculated from unified line broadening theory
of \citet{anderson1952} and a Fermi interaction potential.
Evidence was presented for the non-linear dependence of line width
on density  when the impact approximation is not valid.
Until this pioneering work the vast majority of experimental studies of
the spectral line broadening of alkakies were undertaken at low rare gas perturber
density \citep{allard1982} that does not allow us to distinguish multiple perturber
effects.
As pointed out by \citet{allard1980}, the discontinuity
in width versus \mbox{alkali} series member should arise when the satellite suddenly becomes
important in determining the total width. Such effects might also be expected in
cesium spectra, but published studies have been concerned primarily with
density dependence of the shift and width, rather than the variation
through the series \citep{gilbert1980}.
The discontinuity for thallium does occur in the density dependence
as well, for $11s$ at about $3\times10^{18}$~cm$^{-3}$, as is apparent in
Fig.~9 of \citet{kielkopf1983}. 
Generally in conclusion,  the higher the degree of excitation the lower the perturber
density required to show these discontinuities of the line parameters.

\section*{Data availability}
Complete K--He/H$_2$ opacity tables for the $D1$ and $D2$ components of the $4p-5s$ transitions will be available at CDS anonymous FTP archive for $T=1250$~K.

\section*{Acknowledgements} 
We thank the referee for valuable comments that helped improve the manuscript.  This work was supported by CNES, as part of the French
  contribution to the Ariel Space Mission.

\bibliographystyle{aa} % style aa.bst
%\bibliography{K4p5s}
%\begin{thebibliography}{34}

%

\end{document}